\documentclass[12pt]{article}
\usepackage[a4paper]{geometry}
\usepackage[italian,english]{babel}
\usepackage[T1]{fontenc}

\usepackage{amsmath, accents}
\usepackage{amsfonts}
\usepackage{amstext}
\usepackage{amssymb}
\usepackage{amsthm}
\usepackage{amscd}
\usepackage{mathrsfs}
\usepackage{bbold}
\usepackage{dsfont}
\usepackage{bbm}

\usepackage[pagebackref,draft=false]{hyperref}
\hypersetup{colorlinks,
linkcolor=myrefcolor,
citecolor=mycitecolor,
urlcolor=myurlcolor}

\usepackage[capitalize]{cleveref}
\usepackage{cite}
\usepackage{caption}
\usepackage{etaremune}

\usepackage{xcolor}
\definecolor{myurlcolor}{rgb}{0,0,0.4}
\definecolor{mycitecolor}{rgb}{0,0.5,0}
\definecolor{myrefcolor}{rgb}{0.5,0,0}
\usepackage{graphicx}
\usepackage{tikz}
\usepackage{tikz-cd}
\usetikzlibrary{graphs,decorations.pathmorphing,decorations.markings}

\usepackage{verbatim}
\usepackage{lipsum}
\usepackage{etoolbox}
\usepackage{makeidx}
\usepackage{sectsty}
\usepackage{dsfont}
\usepackage{enumitem} 
\usepackage[]{latexsym}
\usepackage{braket}
\usepackage{caption}
\usepackage[utf8]{inputenx}
\usepackage{lmodern}
\usepackage{textcomp}
\usepackage{microtype}
\usepackage{totcount}
\usepackage{blindtext}

\newtheorem{theorem}{Theorem}[section]
\newtheorem{remark}[theorem]{Remark}

\newtheorem{formulation}{Formulation}
\newtheorem*{proof*}{Proof}

\newcommand{\be}{\begin{equation}}
\newcommand{\ee}{\end{equation}}
\newcommand{\bea}{\begin{eqnarray}}
\newcommand{\eea}{\end{eqnarray}}
\newcommand{\vsp}{\vspace{0.4cm}}

\newcommand{\ac}{\mathscr{S}}

\newcommand{\lag}{\mathscr{L}}

\newcommand{\dd}{{\rm d}}
\newcommand{\de}{\partial}

\title{The inverse problem within free Electrodynamics and the coisotropic embedding theorem}

\author{L. Schiavone$^{1,2}$  \href{https://orcid.org/0000-0002-1817-5752}{\includegraphics[scale=0.7]{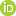}} \\
\footnotesize{\footnotesize{$^{1}$\textit{Departamento de Matem\'aticas, Universidad Carlos III de Madrid, Legan\'es, Madrid, Spain}}} \\
\footnotesize{$^{2}$\textit{e-mail: \texttt{lschiavo[at]math.uc3m.es}}} 
}

\begin{document}

\maketitle
\tableofcontents

\begin{abstract}

We present the coisotropic embedding theorem as a tool to provide a solution for the inverse problem of the calculus of variations for a particular class of implicit differential equations, namely the equations of motion of free Electrodynamics.
%%%
\end{abstract}

%%%%%%%%%%%%%%%%%%%%%%%%%%%%%%%%%%%%%%%%%%%%%%%%%%%%%%%%%%%%%%%%%%%%%%%%%%%%%%%%%%%%%%%%%%%%%%

\section{Introduction}
\label{Sec:Introduction}
% \addcontentsline{toc}{section}{\nameref{Sec:Introduction}}

The Inverse Problem of the Calculus of Variations is an old problem of Mathematics first considered by Helmholtz \cite{Helmholtz1887-Inverse_Problem} in the end of $19^{th}$ century and Douglas \cite{Douglas1941-Inverse_Problem} a few decades later.
%%%

\noindent Taking into account the huge amount of literature appeared after \textit{Helmholtz} and \textit{Douglas}, it is possible to recognize at least two formulations of the inverse problem of the calculus of variations.

\begin{formulation} 
\label{Formulation 1 inverse problem}
In the most general formulation, one starts with a set $\mathscr{A}$ of curves over a manifold $\mathcal{M}$, representing, for instance, experimental data of some physical experiment, and asks whether there exists some function $\ac \,:\; \Gamma(\mathcal{M}) \to \mathbb{R}$, $\Gamma(\mathcal{M})$ denoting the space of curves over $\mathcal{M}$, such that:
    \be
    \delta \ac_\gamma \,=\, 0 \,\, \iff \,\, \gamma \in \mathscr{A} \,,
    \ee
$\delta \ac$ denoting the variation of $\ac$ along any direction \cite{Marmo-Saletan-Sim-Vit1985-Dynamical_Systems, Bozis1995-Inverse_Problem_dynamics, Sarlet-Mestdag-Prince2013-Szebehely_Inverse_Problem, Sarlet-Mestdag-Prince2017-Szebehely_Inverse_Problem_dim3, Ciaglia-Marmo-Schiavone2019-From_trajectories_to_commutation_relations}.
%%%
Essentially, given a set of curves over some manifold, one asks whether they may be considered as extrema of a suitable functional.
%%%
\end{formulation}

\begin{formulation}
\label{Formulation 2 inverse problem}
A slightly less general formulation starts with the assumption that one already knows that elements of $\mathscr{A}$ are solutions of a set of differential equations, say $\mathscr{D}$\footnote{Let us stress that the step of constructing $\mathscr{D}$ from $\mathscr{A}$ is far from being trivial. It is indeed, the core of the work of a theoretical physicists and it is maybe a hopeless problem that of thinking of an algorithmic procedure to do it \cite{Marmo-Saletan-Sim-Vit1985-Dynamical_Systems, Ciaglia-Marmo-Schiavone2019-From_trajectories_to_commutation_relations}.}.
%%%
In this case the inverse problem of the calculus of variation is stated as the search for a Lagrangian function whose Euler-Lagrange equations coincide with the system $\mathscr{D}$ (here there is a huge amount of literature, see, for instance \cite{Santilli1978-Foundations_Theoretical_Mechanics, Henneaux1982-Inverse_Problem, Henneaux1984-Inverse_Problem_Fields, Krupka2015-Variational_Geometry} and references therein).
%%%
\end{formulation}

\begin{remark}
It is worth mentioning that there exists at least another way the inverse problem is stated within theoretical physics literature, namely, the so called inverse problem for Poisson dynamics.
%%%
It is stated as follows: given a set of curves $\mathscr{A}$ over a manifold $\mathcal{M}$, is there an Hamiltonian function $H$ on $\mathcal{M}$ and a Poisson bivector field $\Lambda$ such that elements of $\mathscr{A}$ are integral curves of the Hamiltonian vector field $X_H \,=\, \Lambda(\dd H)$? \cite{Carinena-Ibort-Marmo-Stern1995-Inverse_problem_Poisson}
%%%
We will not consider this kind of problem within this manuscript where we will focus on the Lagrangian side of the story.
%%%
\end{remark}

Before discussing the main results available in the literature regarding the inverse problem, let us make a digression about the motivations (both mathematical and physical) behind the interest in its solution. 
%%%

\noindent From the mathematical point of view, providing a Lagrangian formulation for a given set of differential equations may prove to be useful. 
%%%
For instance, it may be possible to apply indirect integration methods to find solutions, or to argue about the regularity of the solutions.
%%%

\noindent From the Physical point of view, it is well-known the direct relationship that a Lagrangian formulation has with symmetries and constants of the motion (see \cite{Marmo-Schiav-Zamp2020-NoetherI, Marmo-Schiav-Zamp2021-Noether2} and references therein), and most of the standard approaches to the quantization of dynamical systems usually starts from a Lagrangian point of view. 
%%%

\noindent Moreover, in the quantum mechanical setting, it was showed \cite{Ciaglia-DC-Fig-Manko-Marmo-Schiav-Vit-Ventri2019-Nonlinear_dynamics} that a Lagrangian formulation provides a tool to algorithmically restrict the dynamics to arbitrary submanifolds of the carrier manifold where the dynamics is defined.
%%%
This may be of interest both from the experimental point of view (in case only a class of states can be actually prepared in the laboratory) and from the computational point of view (in some cases the equations of motion may be not analytically integrable and an approximation inside a class of trial states may simplify computations). 
%%%

\noindent Finally, a Lagrangian formulation of the dynamics proved to be of interest also within the context of Information Theory.
%%%
Following the approach due to \textit{Amari} \cite{Amari2016-Information_geometry}, a fundamental ingredient to construct metrics on spaces of probability distributions are the so called \textit{divergence functions} (an intrinsic approach to Amari procedure can be found in \cite{Ciaglia-DC-Laud-Marmo-Mele-Ventr-Vit2018-Pedagogical}). 
%%%
Within this context it was shown in \cite{Ciaglia-DC-Felice-Manc-Marmo-PP2017-Potential_functions, Chirico-Malago-Pistone2022-Statistical_bundle} that a suggestive interpretation of the action functional as a divergence function exists. 
%%%

In our review of the main known results about the inverse problem, we will first consider the general setting of partial differential equations.
%%%

\paragraph{Partial differential equations}

The state of the art about the inverse problem for partial differential equations of any order is elegantly resumed in \cite{Krupka2015-Variational_Geometry}.
%%%
It is also worth mentioning some previous references that played a major role in the development of the general theory resumed in  \cite{Krupka2015-Variational_Geometry}, such as \cite{Henneaux1984-Inverse_Problem_Fields, Anderson-Duchamp1984-Variational_principles} and references therein.
%%%

\noindent In \cite{Krupka2015-Variational_Geometry} the author considers general systems of partial differential equations of order $k$ of the type:
\be \label{Eq:k-order pde}
E_\sigma(x^j,\, y^\sigma,\, y^\sigma_{j_1},\, y^\sigma_{j_1 j_2},\, ...,\, y^\sigma_{j_1...j_k}) \,=\, 0 \,,
\ee
where $x^j$ ($j=1,..., n$) are independent variables, $y^\sigma$ ($\sigma=1,...,l$) are the dependent variables, $y^\sigma_{j_1...j_k}$ represent the derivatives of order $k$ of the dependent variables with respect to the independent ones and $E_\sigma$ are smooth functions.
%%%
The set of independent and dependent variables (together with the derivatives up to order $k$) can be considered as a set of fibered coordinates over the $k$-order jet bundle $\mathbf{J}^k \pi$ of the fibration $\pi \, :\; Y \to X$ where $\left\{\,x^j,\, y^\sigma \right\}_{j=1,...,n;\,\sigma=1,...,l}$ are fibered coordinates on $Y$ associated to the local coordinates $\left\{\, x^j \,\right\}_{j=1,...,n}$ on $X$ and $\pi$ is the projection onto the first factor.
%%%
On the other hand, the functions $E_\sigma$ can be considered as coefficients of a $(n+1)$-form on $\mathbf{J}^k \pi$ of the type:
\be
E \,=\, E_\sigma \eta^\sigma \wedge \underbrace{\dd x^1 \wedge ... \wedge \dd x^n}_{vol_X} \,,
\ee
called a \textit{source form}, where $\eta^\sigma \,=\, \dd y^\sigma - y^\sigma_j \dd x^j$.
%%%
In this setting, a Lagrangian is a $n$-form on $\mathbf{J}^k \pi$ of the type:
\be
\lambda \,=\, \lag(x^j,\, y^\sigma,\, y^\sigma_{j_1},\, y^\sigma_{j_1 j_2},\, ...,\, y^\sigma_{j_1...j_k}) \, vol_X \,,
\ee
where we are assuming $X$ to have a volume form that we are denoting by $vol_X$.
%%%
Its Euler-Lagrange equations can be written as coefficients of a source form of the type mentioned above:
\be
E^\lag \,=\, E^\lag_\sigma \eta^\sigma \wedge vol_X \,:=\, \left[\, \frac{\de \lag}{\de y^\sigma} +  \sum_{m=1}^k (-1)^m d_{j_1}...d_{j_m} \frac{\de \lag}{\de y^\sigma_{j_1...j_m}} \,\right] \eta^\sigma \wedge vol_X \,,
\ee
where
\be
d_{j} \,=\, \frac{\de}{\de x^j} + \sum_{j_1 \leq ... \leq j_k} y^\sigma_{j_1...j_k j} \frac{\de }{\de y^\sigma_{j_1...j_k}}  \,, 
\ee
is the so-called \textit{total derivative operator}.
%%%
With this in mind, the set of partial differential equations is said to be \textit{variational} if:
\be
E \,=\, E^\lag \,,
\ee
for some $\lag$.
%%%
\begin{remark}
    This definition can be made more general by considering a proportionality of the type
    \be
    E_\sigma \,=\, A_\sigma^\mu E^\lag_\mu \,,
    \ee
    where $A_\sigma^\mu$ is referred to as an integrating factor and is any matrix such that $E^\lag_\sigma \eta^\sigma \wedge vol_X$ and $A_\sigma^\mu E^\lag_\mu \eta^\sigma \wedge vol_X$ have the same set of zeroes.
    %%%
    This instance is not considered in \cite{Krupka2015-Variational_Geometry}.
    %%%
    It is considered in \cite{Henneaux1984-Inverse_Problem_Fields, Saunders2010-30Years_Inverse_Problem} but only for second order equations.
    %%%
\end{remark}
%%%

\noindent The main result contained in \cite[Thm $13$]{Krupka2015-Variational_Geometry} is the following.
%%%
\begin{theorem} \label{Thm:locally variational}
    $E$ is variational iff:
    \be
    H_{\sigma \mu}^{j_1...j_m}(E) \,=\, 0 \,,
    \ee
    where:
    \be
    H_{\sigma \mu}^{j_1...j_m}(E) \,=\, \frac{\de E_\sigma}{\de y^\mu_{j_1...j_m}} - (-1)^m \frac{\de E_\mu}{\de y^\sigma_{j_1...j_m}} - \sum_{r=m+1}^k (-1)^r {r\choose m} d_{p_{m+1}}... d_{p_{r}} \frac{\de E_\mu}{\de y^\sigma_{j_1...j_m p_{m+1}...p_r}} \,,
    \ee
    which can be seen as the coefficients of a $(m+2)$-form on $\mathbf{J}^k \pi$ of the type:
    \be
    H(E) \,=\, \frac{1}{2} \sum_{m=0}^k H^{j_1...j_m}_{\sigma \mu}(E) \eta^\mu_{j_1...j_m} \wedge \eta^\sigma \wedge vol_X \,,
    \ee
     where $\eta^\mu_{j_1...j_m} \,=\, \dd y^\mu_{j_1...j_m} - y^\mu_{j_1...j_m j} \dd x^j$, usually referred to as Helmholtz form.
    %%%
    The coefficients $H_{\sigma \mu}^{j_1...j_m}(E)$ are usually referred to as Helmholtz expressions of E.
    %%%
\end{theorem}
%%%
The theorem above says that a partial differential equation of the type \eqref{Eq:k-order pde} is variational iff its Helmholtz expressions vanish.
%%%
It is worth stressing that the differential forms $\lambda$, $E^\lag$ and $H(E)$ are only locally defined on $\mathbf{J}^k \pi$ in general.
%%%
With this in mind Thm. \ref{Thm:locally variational} gives necessary and sufficient conditions for a partial differential equation to be \textit{locally variational}.
%%%
The possibility of globally defining the differential forms above is related to a cohomological obstruction.
%%%
In particular, if the $(n+1)$ DeRham cohomology group $H^{n+1} Y$ vanishes, then $\lambda$, $E^\lag$ and $H(E)$ are globally defined and $E$ is said to be \textit{globally variational} \cite[Thm $13$]{Krupka2015-Variational_Geometry} (see also \cite{Anderson-Duchamp1980-Global_Variational_Principles, Tulczyjew1977-Lagrange_complex} for a global approach to the inverse problem).
%%%

\begin{remark} \label{Rem:Tonti-Veinberg Lagrangian}
    Note that the theorems above are not constructive, in the sense that they do not provide an explicit form for $\lambda$. 
    %%%
    It is worth noting that the explicit form of $\lambda$ is not unique in general, and a possible Lagrangian providing a solution for the inverse problem when the hypothesis of Thm. \ref{Thm:locally variational} are met is the so-called Tonti-Veinberg Lagrangian \cite{Tonti1972-Variational, Vainberg1964-Variational_Methods}:
    \be
    \lag \,=\, y^\sigma \int_0^1 E_\sigma (x,\, ty^\sigma,\, t y^\sigma_{j_1}, ..., t y^\sigma_{j_1...j_k} ) \dd t \,.
    \ee
\end{remark}

\vsp 

We believe there is another point of view from which to look at the inverse problem that stems from a different observation regarding the initial mathematical objects used to model physical reality.
%%%
Instead of focusing on forms on jet bundles, we start from the fact that the results of physical experiments can be usually collected into a set of curves over some configuration space that we assume to be a manifold (an assumption that is often met in practice).
%%%
In particular, such a manifold is finite-dimensional within mechanical systems and is usually referred to as \textit{configuration space}.
%%%
Within field theory such a manifold is infinite-dimensional and represents the space of fields at each instant of time.
%%%
With this in mind, curves over this infinite-dimensional configuration space represent a field theoretic generalization of the concept of trajectory, i.e., time evolution of the fields of the theory.
%%%

\noindent Our point of view starts from the observation that the most direct relation between a set of curves $\mathscr{A}$ over a manifold, $\mathcal{M}$, and differential equations is given, both in the finite and infinite-dimensional case, by vector fields on $\mathcal{M}$.
%%%
Indeed, given $\mathscr{A}$, a natural thing one may do is to try to construct a vector field $\Gamma$ on $\mathcal{M}$ being everywhere tangent to the elements of $\mathscr{A}$.
%%%
If this is possible, recalling that to each vector field on a manifold it is associated a set of ordinary differential equations, then one can claim that the set of ordinary differential equations defining $\Gamma$ has $\mathscr{A}$ as space of solutions.
%%%
Of course this is not possible for any given set $\mathscr{A}$ of curves, even if there are many examples where this procedure succeeds, a famous, finite-dimensional, one being represented by the \textit{Kepler's problem} where \textit{Kepler} reconstructed the gravitational field generated by the sun observing planet's orbits (see \cite{Marmo-Saletan-Sim-Vit1985-Dynamical_Systems}).
%%%
There are various results available in the literature about the inverse problem for systems of ordinary differential equations that take advantage of such a formulation in terms of vector fields and that we report below.
%%%

\paragraph{Ordinary differential equations}

The main known results about the inverse problem for sets of ordinary differential equations regard two classes of equations: 
\begin{itemize}
    \item Second order differential equations (\textsc{SODE}) on some manifold $\mathcal{M}$, namely, when $\mathscr{D}$ is of the type:
    \be
    \ddot{q}^j(t) \,=\, f^j(q(t),\, \dot{q}(t)) \,,
    \ee
    where $\{\,q^j\,\}_{j=1,...,\mathrm{dim}\mathcal{M}}$ denotes a system of local coordinates on $\mathcal{M}$.
    %%%
    In this case $\mathscr{A}$ is the set of (the projections onto $\mathcal{M}$ of) integral curves of the vector field:
    \be
    \Gamma \,=\, v^j \frac{\de}{\de q^j} + f^j(q,\, v) \frac{\de}{\de v^j} \,,
    \ee
    on $\mathbf{T}\mathcal{M}$, usually referred to as \textit{second order vector field}.
    %%%
    Above $\{\, q^j,\, v^j \,\}_{j=1,...,\mathrm{dim}\mathcal{M}}$ denotes a system of local coordinates on $\mathbf{T}\mathcal{M}$.
    %%%
    Results about the solution for the inverse inverse problem in this case can be resumed in the following theorem \cite{Morandi-Ferrario-LoVecchio-Marmo-Rubano1990-Inverse_Problem, Crampin1981-Inverse_Problem}.
    \begin{theorem} \label{Thm:inverse problem sode}
        Given a second order vector field $\Gamma$ over the tangent bundle of a manifold, $\mathbf{T}\mathcal{M}$, if there exists a symplectic form $\omega$ on $\mathbf{T}\mathcal{M}$ such that:
        \be
        \mathcal{L}_\Gamma \omega \,=\, 0 \,,
        \ee
        and
        \be
        \omega(V,\, W) \,=\, 0 \,, \;\;\;\; \forall \,\, V,\, W \in \mathfrak{X}^v(\mathbf{T}\mathcal{M}) \,,
        \ee
        where $\mathfrak{X}^v(\mathbf{T}\mathcal{M})$ denotes the space of vertical vector fields on $\mathbf{T}\mathcal{M}$, then there exists a local Lagrangian function $\lag$ on $\mathbf{T}\mathcal{M}$ such that:
        \be
        \omega_\lag \,:=\, - \dd \dd_S \lag \,=\, -\dd \circ  S (\dd \lag) \, \equiv \omega \;\;\;\;\; \text{where} \;\; S \,=\, \dd q^j \otimes \frac{\de}{\de v^j} \,,
        \ee
        and 
        \be
        i_\Gamma \omega_\lag \,=\, \dd E_\lag  \;\;\;\;\; \text{where} \;\; E_\lag \,=\, \Delta(\lag) - \lag \;\;\; \text{and} \;\; \Delta \,=\, v^j \frac{\de}{\de v^j} \,.
        \ee
    \end{theorem}
    %%%
    The analogue results for non-autonomous mechanical systems can be found, for instance, in \cite{Krupkova1997-Geometry_OVE, Krupkova-Prince2008-Second_Order_ODE_Inverse_Problem, Crampin-Prince-Thompson1984-Helmholtz_Conditions_time_dependent}.
    \begin{remark}
        Note that also in this case the theorem is not constructive and does not provide an explicit form for $\lag$ which may be also not unique.
    \end{remark}
    %%%
    \item First order differential equations (\textsc{FODE}) on some manifold $\mathcal{M}$, namely when $\mathscr{D}$ is of the type:
    \be
    \dot{q}^j(t) \,=\, f^j(q(t)) \,,
    \ee
    where $\left\{\, q^j \,\right\}_{j=1,...,\mathrm{dim}\mathcal{M}}$ denotes a system of local coordinates on $\mathcal{M}$.
    %%%
    In this case $\mathscr{A}$ is the set of integral curves of the vector field:
    \be
    \Gamma \,=\, f^j(q) \frac{\de}{\de q^j} \,,
    \ee
    on $\mathcal{M}$.
    %%%
    Results about the solution of the inverse problem in this case can be resumed in the following theorem \cite{Ibort-Solano1991-Inverse_Problem_coupled_systems} (see also \cite{Ciglia-DC-Ibort-Marmo-Schiav-Zamp2020-Heisenberg}).
    %%%
    \begin{theorem} \label{Thm:inverse problem fode}
        Given a vector field $\Gamma$ over a manifold $\mathcal{M}$, if there exists a symplectic form $\omega$ on $\mathcal{M}$ such that:
        \be
        \mathcal{L}_\Gamma \omega \,=\, 0 \,,
        \ee
        then, there exists a local Lagrangian function $\lag$ on $\mathbf{T}\mathcal{M}$ such that the pre-symplectic Hamiltonian system $(\mathbf{T}\mathcal{M},\, \omega_\lag,\, E_\lag)$ gives, by applying the pre-symplectic constraint algorithm (see \cite{Gotay-Nester1978-Presymplectic_DiracBergmann} for the definition of the algorithm), that:
        \begin{itemize}
            \item $\mathcal{M}_\infty \equiv \mathcal{M}$ is the final stable manifold of the algorithm;
            \item ${\omega_\lag}_\infty \,=\, \mathfrak{i}_\infty^\star \omega_\lag \equiv \omega$, where $\mathfrak{i}_\infty$ is the immersion of $\mathcal{M}_\infty \,\equiv \, \mathcal{M}$ into the original manifold $\mathbf{T}\mathcal{M}$;
            \item $\Gamma$ satisfies $i_\Gamma {\omega_\lag}_\infty \,=\, i_\Gamma \omega \,=\, \dd {E_\lag}_\infty$ where ${E_\lag}_\infty \,=\, \mathfrak{i}_\infty^\star E_\lag$.
        \end{itemize}
        This means that, given a vector field $\Gamma$ over a manifold, if the manifold is symplectic and if the vector field preserves the symplectic structure, there exists a degenerate Lagrangian on the tangent bundle of the symplectic manifold giving rise to first order Euler-Lagrange equations coinciding with the set of ordinary differential equations defining $\Gamma$.
        %%%
    \end{theorem}
    \begin{remark}
        In this case the theorem is also constructive, since it provides a specific (even if not unique) Lagrangian for $\Gamma$.
        %%%
        In particular, if locally $\omega \,=\, \de_j B_k \dd q^j \wedge \dd q^k$, then the explicit form of $\lag$ reads:
        \be
        \lag \,=\, B_j v^j - E_\lag \,,
        \ee
        where $E_\lag$ is any function on $\mathbf{T}\mathcal{M}$ satisfying ${E_\lag}_\infty \,=\,, \mathfrak{i}_\infty^\star E_\lag$ and where ${E_\lag}_\infty$ is exactly obtained as solution of the equation $i_\Gamma \omega \,=\, \dd {E_\lag}_\infty$ where $\Gamma$ and $\omega$ are known.
        %%%
    \end{remark}
\end{itemize}

\noindent We can now be more precise about our alternative (to that of \textit{Krupka et al.}) approach to the inverse problem, that we will adopt for the case of free Electrodynamics in vacuum in the present manuscript and we plan to extend to non-Abelian gauge theories in future papers.
%%%
Such an approach starts from two fundamental observations:
\begin{itemize}
    \item as we argued in the discussion below Rem. \ref{Rem:Tonti-Veinberg Lagrangian}, very often the results of experimental measures over a physical system can be modeled as curves over a configuration space.
    %%%
    In particular, these may be curves over a finite-dimensional manifold modeling the physical trajectories of a finite-dimensional mechanical system perceived by an observer or, within field theories, curves over an infinite-dimensional space of fields obtained out of subsequent measurements at subsequent instants of time of the fields of the theory (e.g. subsequent measurements of the electric or magnetic field produced by some system of sources).
    %%%
    With this in mind, as we will see in Sec. \ref{Sec:Electrodynamics and pre-symplectic manifolds} for the case of Electrodynamics, equations of motion within gauge theories can be formulated in terms of a vector field over a (infinite-dimensional) pre-symplectic manifold.
    %%%
    Even if it may seem quite abstract, we want to stress once more that this formulation follows naturally from modeling experimental data as curves over a manifold and from the idea (already mentioned below Rem. \ref{Rem:Tonti-Veinberg Lagrangian}) that the most direct relation between curves and differential equations is given by vector fields (tangent, eventually, to the set of curves);
    \item there exists a canonical geometrical construction that relates a pre-symplectic manifold to a (larger) symplectic manifold, i.e. the so-called \textit{coisotropic embedding theorem}.
    %%%
    Such a theorem can be used to map the above mentioned setting, namely a vector field over a pre-symplectic manifold, into the setting of Thm. \ref{Thm:inverse problem fode} and used to solve the inverse problem.
    %%%
\end{itemize}
%%%
Therefore, the goal of the paper is to show that within free Electrodynamics in vacuum, the inverse problem can be solved by using a suitable generalization of Thm. \ref{Thm:inverse problem fode} via a technique related to the coisotropic embedding theorem.
%%%

\vsp

\noindent The paper is organized as follows.
%%%
We will devote Sec. \ref{Sec:Electrodynamics and pre-symplectic manifolds} to show how to formulate the equations of motion of free Electrodynamics in vacuum\footnote{The generalization to non-Abelian gauge theories is mentioned in the Conclusions of the manuscript.} in terms of a family of vector fields over a suitable infinite-dimensional pre-symplectic manifold.
%%%
Then, after reviewing the content of the coisotropic embedding theorem in \ref{Sec:The coisotropic embedding theorem}, we will provide, in Sec. \ref{Sec:The inverse problem for free Electrodynamics}, a generalization of Thm. \ref{Thm:inverse problem fode} to the family of vector fields describing free Electrodynamics.
%%%

%%%%%%%%%%%%%%%%%%%%%%%%%%%%%%%%%%%%%%%%%%%%%%%%%%%%%%%%%%%%%%%%%%%%%%%%%%%%%%%%%%%%%%%%%%%%%%%%%%%%%%%%%%%%%%%%%%%%%%%%%%%%%%%%%%%%%%%%%%%%%%%%%%%%%%%%%%%%%%%%%%%%%%%%%%%%%%%%%%%%%%%%%%%%%%%%%%

\section{Electrodynamics and pre-symplectic manifolds}
\label{Sec:Electrodynamics and pre-symplectic manifolds}

We devote this section to recall how to formulate free Electrodynamics in vacuum in terms of a family of vector fields over an infinite-dimensional pre-symplectic manifold (see also \cite{Ciaglia-DC-Ibort-Marmo-Schiav-Zamp2021-Cov_brackets_toappear}).
%%%

\noindent Given a space-time splitting for the Minkowski space-time $(\mathbb{R}^4,\, g)$ with the system of adapted coordinates $\left\{\, t,\, \underline{x} \,\right\}$, free Electrodynamics in vacuum can be described in terms of a \textit{vector potential} $\underline{A}(t,\, \underline{x})$ and an \textit{electric field} $\underline{E}(t,\, \underline{x})$ obeying \textit{Maxwell's equations}:
\be
\begin{split}
\mathrm{div} \underline{E} \,&=\, 0  \,,\\
\frac{\de \underline{E}}{\de t}  \,&=\, \Delta \underline{A} - \mathrm{grad} \, \mathrm{div} \underline{A} \,, \\
\frac{\de \underline{A}}{\de t}  \,&=\, \underline{E} + \mathrm{grad} \phi \,.
\end{split}
\ee
for any arbitrary function $\phi$ whose appearance in the equations of motion does not affect the value of all physical measurable quantities of the theory.
%%%

\noindent Having in mind the multisymplectic formulation of free Electrodynamics \cite{Ibort-Spivak2017-Covariant_Hamiltonian_YangMills, Ciaglia-DC-Ibort-Marmo-Schiav-Zamp2021-Cov_brackets_toappear}, vector potential and electric field can be interpreted, at each $t$, as the coefficients of a differential $1$-form and a vector field on $\mathbb{R}^3$:
\be
A \,=\, A_j(\underline{x}) \dd x^j \,, \qquad E \,=\, E^j(\underline{x}) \frac{\de}{\de x^j} \,, \;\;\;\;\; j=1,2,3 \,.
\ee
%%%
In particular, in \cite{Ciaglia-DC-Ibort-Marmo-Schiav-Zamp2021-Cov_brackets_toappear} it was shown that a rigorous infinite-dimensional differential geometric approach can be carried on the space of those $A_j$ being $\mathcal{H}^{\frac{3}{2}}$ functions on $\mathbb{R}^3$ and those $E^j$ being $\mathcal{H}^{\frac{1}{2}}$ functions on $\mathbb{R}^3$.
%%%
Denoting by $\mathcal{F} \,=\, \left[\,{\mathcal{H}^{\frac{3}{2}}(\mathbb{R}^3)}\,\right]^3 \times \left[\,{\mathcal{H}^{\frac{1}{2}}(\mathbb{R}^3)}\,\right]^3$ the space of vector potentials and electric fields at each $t$, then solutions to Maxwell's equations are curves on $\mathcal{F}$, say $(A(s),\, E(s))$, with $s \in \mathbb{I} \subset \mathbb{R}$ being the parameter describing the curve, satisfying:
\be \label{Eq:Maxwell}
\begin{split}
\de_k E^k(s) \,&=\, 0 \;\;\; \forall \,\, s \in \mathbb{I} \,, \\
\frac{d}{ds} A_k(s) \,&=\, \delta_{kj} E^j(s) + \de_k \phi \,, \\
\frac{d}{ds} E^k(s) \,&=\, \delta^{kj} \left(\, \Delta A_j(s) - \de_j \delta^{ml} \de_m A_l(s) \,\right) \,.
\end{split}
\ee
%%%
The first equation above, i.e. the so-called \textit{Gauss' constraint}, selects a closed subspace of $\mathcal{F}$ \cite{Ciaglia-DC-Ibort-Marmo-Schiav-Zamp2021-Cov_brackets_toappear}, i.e. $\mathcal{M} \,=\, \left[\,{\mathcal{H}^{\frac{3}{2}}(\mathbb{R}^3)}\,\right]^3 \times \left[\,{\mathcal{H}^{\frac{1}{2}}(\mathrm{div}0,\, \mathbb{R}^3)}\,\right]^3$, where $\left[\,{\mathcal{H}^{\frac{1}{2}}(\mathrm{div}0,\, \mathbb{R}^3)}\,\right]^3$ is the space of $\mathcal{H}^{\frac{1}{2}}$ functions on $\mathbb{R}^3$ with zero divergence. 
%%%
Such a subspace is a pre-symplectic manifold \cite{Ciaglia-DC-Ibort-Marmo-Schiav-Zamp2021-Cov_brackets_toappear}.
%%%
Indeed, the whole manifold $$\left[\, \mathcal{H}^{\frac{3}{2}}(\mathbb{R}^3)\,\right]^3 \times \left[\, \mathcal{H}^{\frac{1}{2}}(\mathbb{R}^3)\,\right]^3$$ has a symplectic structure reading:
\be
\omega \,=\, \delta A_k \wedge \delta E^k \,,
\ee
where, with a slight abuse of notation we are still denoting by $E^k$ an element in the whole $\mathcal{H}^{\frac{1}{2}}(\mathbb{R}^3)$.
%%%
With the notation above we mean that, given two tangent vectors of the type $$\mathbb{X}_{(A,E)} \,:=\, \left(\, {\left(\mathbb{X}_{(A,E)}\right)_{A_k}},\, {\left(\mathbb{X}_{(A,E)}\right)_{E^k}} \,\right) \in \mathbf{T}_{(A,E)} \mathcal{M} \,\simeq\, \mathcal{M} \,=\, \left[\, \mathcal{H}^{\frac{3}{2}}(\mathbb{R}^3)\,\right]^3 \times \left[\, \mathcal{H}^{\frac{1}{2}}(\mathbb{R}^3)\,\right]^3 \,,$$ at $(A,\,E) \in \mathcal{M}$, the action of $\omega$ at the point $(A,\,E)$ on $\mathbb{X}_{(A,E)}$ and $\mathbb{Y}_{(A,E)}$ reads:
\be
\begin{split}
\omega_{(A,E)}\left(\, \mathbb{X}_{(A,E)},\, \mathbb{Y}_{(A,E)} \,\right) \,=\, \int_{\mathbb{R}^3} \biggl[\, &{\left(\mathbb{X}_{(A,E)}\right)_{A_k}}(x) {\left(\mathbb{Y}_{(A,E)}\right)_{E^k}}(x) \\  
- &{\left(\mathbb{Y}_{(A,E)}\right)_{A_k}}(x) {\left(\mathbb{X}_{(A,E)}\right)_{E^k}}(x) \,\biggr] \, \dd^3 x \,,
\end{split}
\ee
where the integral is well defined because of the regularity conditions chosen.
%%%
The $2$-form $\omega$ is evidently closed and non-degenerate.
%%%
Having in mind the following decomposition of the Hilbert space $\left[\, \mathcal{H}^{\frac{3}{2}}(\mathbb{R}^3) \,\right]^3$ \cite{Lions1990-vol3, Ciaglia-DC-Ibort-Marmo-Schiav-Zamp2021-Cov_brackets_toappear}:
\be
\left[\, \mathcal{H}^{\frac{3}{2}}(\mathbb{R}^3) \,\right]^3 \,=\, \left[\, \mathcal{H}^{\frac{3}{2}}(\mathrm{div}0,\, \mathbb{R}^3) \,\right]^3 \oplus \mathrm{grad} \mathcal{H}^{\frac{5}{2}}(\mathbb{R}^3) \,,
\ee
where $\mathrm{grad} \mathcal{H}^{\frac{5}{2}}(\mathbb{R}^3)$ is the space of $\mathcal{H}^{\frac{3}{2}}$ functions being the gradient of a $\mathcal{H}^{\frac{5}{2}}$ function and whose points will be accordingly denoted by $(\tilde{A},\, \psi)$, the restriction of the symplectic form above to $\mathcal{M}$ (that we still call $\omega$ with an abuse of notation) reads:
\be
\omega \,=\, \delta \tilde{A}_k \wedge \delta E^k \,,
\ee
where now, $E^k$ are denoting the components of an element of $\left[\,\mathcal{H}^{\frac{1}{2}}(\mathrm{div}0,\, \mathbb{R}^3)\,\right]^3$.
%%%
Because of the Gauss' constraint, it has an infinite-dimensional kernel given by:
\be
\mathrm{ker} \omega \,=\, <\left\{\, \de_k \phi \frac{\delta}{\delta \psi_k} \,\right\}> \,\simeq \, \mathrm{grad}\mathcal{H}^{\frac{5}{2}}(\mathbb{R}^3) \,.
\ee
%%%

\noindent On the other hand, the latter equations in \eqref{Eq:Maxwell} are ordinary differential equations on the infinite-dimensional manifold $\mathcal{M}$ defining the components of a vector field.
%%%
Such a vector field is parametrized by the arbitrary function $\phi \in \mathcal{H}^{\frac{5}{2}}(\mathbb{R}^3)$\footnote{This choice of regularity is made in order for its derivatives to be $\mathcal{H}^{\frac{3}{2}}$ functions like the components of $A$.}, thus, the equations above define actually a whole family of vector fields on $\mathcal{M}$ that we will denote by $\Gamma_\phi$.
%%%

%%%%%%%%%%%%%%%%%%%%%%%%%%%%%%%%%%%%%%%%%%%%%%%%%%%%%%%%%%%%%%%%%%%%%%%%%%%%%%%%%%%%%%%%%%%%%%%%%%%%%%%%%%%%%%%%%%%%%%%%%%%%%%%%%%%%%%%%%%%%%%%%%%%%%%%%%%%%%%%%%%%%%%%%%%%%%%%%%%%%%%%%%%%%%%%%%%

\section{The coisotropic embedding theorem}
\label{Sec:The coisotropic embedding theorem}

In this section, we recall the content of the coisotropic embedding theorem that is used in Sec. \ref{Sec:The inverse problem for free Electrodynamics} to prove the main result of this work.
%%%
The theorem was first proved by \textit{M. Gotay} in \cite{Gotay1982-Coisotropic_embedding}.
%%%
We also refer to \cite{Guillemin-Sternberg1990-Symplectic_techniques} for an equivariant version of the theorem.
%%%
Here, we report a slightly more modern version of the theorem due to \textit{Oh} and \textit{Park} \cite{Oh-Park2005-Embedding_coisotropo} that we already used in \cite{Ciaglia-DC-Ibort-Mar-Schiav-Zamp2022-Non_abelian, Ciaglia-DC-Ibort-Mar-Schiav-Zamp2022-Palatini} to construct Poisson brackets on the space of solutions of the equations of motion within non-Abelian gauge theories and in \cite{Ciaglia-DC-Ibort-Marmo-Schiav-Zamp-2022-Symmetry} to construct a one-to-one correspondence between symmetries and constants of the motion within pre-symplectic Hamiltonian systems.
%%%

\begin{theorem}[\textsc{Coisotropic embedding theorem}]
Consider a pre-symplectic smooth Banach manifold $(\mathcal{M},\, \omega)$.
%%%
Denote by $K_m$ the kernel of $\omega$ at each $m \in \mathcal{M}$.
%%%
Assume that, at each $m \in \mathcal{M}$, a complement of $K_m$ into $\mathbf{T}_m \mathcal{M}$, say $\mathbf{T}_m\mathcal{M} \,=\, K_m \oplus W_m$, exists and is given by the projection:
\be
P \;\; :\;\; \mathbf{T}\mathcal{M} \to K \,.
\ee
%%%
Denote by $\pi \;:\;\; \mathbf{K} \to \mathcal{M}$ the so-called characteristic bundle over $\mathcal{M}$, namely, the vector bundle over $\mathcal{M}$ whose fibre at each $m \in \mathcal{M}$ is given by $K_m$.
%%%
Denote by $\tau \;:\;\; \mathbf{K}^\star \to \mathcal{M}$ its dual bundle.
%%%
Then, there exists a neighborhood of the zero section of $\tau$, say $\tilde{\mathcal{M}}$, in which the following symplectic form can be canonically defined:
\be
\tilde{\omega} \,=\, \tau^\star \omega + \dd \vartheta^P \,,
\ee
where:
\be
\vartheta^P_\beta (X) \,=\, \langle \, \beta,\, P \circ T\tau (X) \,\rangle \,,
\ee
with $\beta \in \mathbf{K}^\star$ and $X \in \mathbf{T}_\beta \mathbf{K}^\star$ and where $T\tau$ is the tangent map to $\tau$:
\be
T\tau \;\; :\;\; \mathbf{T}\mathbf{K}^\star \to \mathbf{T}\mathcal{M} \,.
\ee
%%%
The smooth Banach manifold $(\tilde{\mathcal{M}},\, \tilde{\omega})$ is a symplectic manifold (sometimes referred to as symplectic thickening of $(\mathcal{M},\, \omega)$) of which $(\mathcal{M},\, \omega)$ is a coisotropic submanifold, i.e., $\tau^\star \tilde{\omega} \,=\, \omega$.
%%%
\end{theorem}
%%%
A local expression for $\tilde{\omega}$ that will be useful in the sequel can be given by providing a local expression for the projection $P$ in terms of an idempotent $(1,\,1)$-tensor field on $\mathcal{M}$:
\be
P \,=\, P^j \otimes V_j \,,
\ee
where $\left\{\, V_j \,\right\}_{j=1,..., \mathrm{dim}\mathbf{K}}$ is, at each point, a system of generators of $K_m$ and $\left\{\, P^j \,\right\}_{j=1,...,\mathrm{dim}\mathbf{K}}$ is a collection of $1$-forms on $\mathcal{M}$ whose action upon a vector field $Y$ provides its components along $K$ in the chosen basis:
\be
P(Y) \,=\, P^j(Y) V_j \,=:\, {Y^v}^j V_j \,. 
\ee
%%%
Let us denote by $\left\{\, \mu_j \,\right\}_{j=1,...,\mathrm{dim}\mathbf{K}^\star}$ the adapted system of coordinates on $K^\star_m$ for which:
\be
\dd \mu_j \wedge P^j (X,\, Y) \,=\, \langle\, {X_\mu}_j,\, {Y^v}^j \,\rangle - \langle\, {Y_\mu}_j,\, {X^v}^j \,\rangle \,,
\ee
where $X$ and $Y$ are vector fields on $\tilde{\mathcal{M}}$, ${X_\mu}_j$ denotes the components of $X$ along $K^\star_m$ and $\langle \,\cdot\,,\,\cdot\, \rangle$ is the pairing between $K_m$ and its dual.
%%%
With this choice of local coordinates, $\tilde{\omega}$ looks like:
\be
\tilde{\omega} \,=\, \tau^\star \omega + \dd \mu_j \wedge P^j + \mu_j \dd P^j \,.
\ee
%%%

%%%%%%%%%%%%%%%%%%%%%%%%%%%%%%%%%%%%%%%%%%%%%%%%%%%%%%%%%%%%%%%%%%%%%%%%%%%%%%%%%%%%%%%%%%%%%%%%%%%%%%%%%%%%%%%%%%%%%%%%%%%%%%%%%%%%%%%%%%%%%%%%%%%%%%%%%%%%%%%%%%%%%%%%%%%%%%%%%%%%%%%%%%%%%%%%%%

\section{The inverse problem for free Electrodynamics}
\label{Sec:The inverse problem for free Electrodynamics}

In the introduction to this manuscript we recalled how for those dynamical systems whose evolution is described by means of a vector field over a symplectic manifold, it is possible to provide a Lagrangian description for the dynamical system if the symplectic structure is preserved (Thm. \ref{Thm:inverse problem fode}).
%%%
In Sec. \ref{Sec:Electrodynamics and pre-symplectic manifolds} we saw how the equations of motion of free Electrodynamics in vacuum can be formulated in terms of a family of vector fields over a pre-symplectic manifold.
%%%
On the other hand, in Sec. \ref{Sec:The coisotropic embedding theorem}, we recalled how the coisotropic embedding theorem provide a (larger) canonical symplectic manifold starting from a pre-symplectic one, its symplectic thickening.
%%%
In this section we will show how such a symplectic thickening provides a natural framework to transform the setting of Sec. \ref{Sec:Electrodynamics and pre-symplectic manifolds} into the setting of Thm. \ref{Thm:inverse problem fode} and, consequently, how it provides a natural manifold on which one can carry on a Lagrangian formulation of free Electrodynamics.
%%%

\noindent Recall that we formulated it in terms of a family of vector fields:
\be
\Gamma_\phi \,=\, \left(\, \delta_{kj} E^j + \de_k \phi \, \right) \frac{\delta}{\delta A_k} + \delta^{kj} \left(\, \Delta A_j - \de_j \delta^{ml} \de_m A_l \,\right) \frac{\delta}{\delta E^k} \,,
\ee
over the pre-symplectic manifold $(\mathcal{M},\, \omega)$, where:
\be
\mathcal{M} \,=\, \left[\,{\mathcal{H}^{\frac{3}{2}}(\mathbb{R}^3)}\,\right]^3 \times \left[\,{\mathcal{H}^{\frac{1}{2}}(\mathrm{div}0,\, \mathbb{R}^3)}\,\right]^3 \,,
\ee
and:
\be
\omega \,=\, \delta \tilde{A}_k \wedge \delta E^k \,.
\ee
%%%
Recall also that the kernel of $\omega$ reads:
\be
\mathrm{ker} \omega \,=:\, K \,=\, <\left\{\, \de_k \phi \frac{\delta}{\delta A_k} \,\right\}> \,\simeq \, \mathrm{grad}\mathcal{H}^{\frac{5}{2}}(\mathbb{R}^3) \,.
\ee
%%%
Its dual reads:
\be
K^\star \,=\, \left[\,\mathrm{grad}\mathcal{H}^{\frac{5}{2}}(\mathbb{R}^3)\,\right]^\star \,=\, \mathrm{grad}\mathcal{H}^{\frac{5}{2}}(\mathbb{R}^3) \,,
\ee
since, as it is proved in \cite{Ciaglia-DC-Ibort-Marmo-Schiav-Zamp2021-Cov_brackets_toappear} $\mathrm{grad}\mathcal{H}^{\frac{5}{2}}(\mathbb{R}^3)$ is an Hilbert space itself being $\mathrm{grad}$ a closed operator on $\mathcal{H}^{\frac{5}{2}}(\mathbb{R}^3)$. 
%%%
We will denote by $\{\,\mu^k\,\}_{k=1,2,3}$ a system of coordinates on $K^\star$.
%%%
Consequently:
\be
\tilde{\mathcal{M}} \,=\, \mathcal{M} \times \mathrm{grad}\mathcal{H}^{\frac{5}{2}}(\mathbb{R}^3) \,.
\ee
%%%
\noindent Recalling the content of Sec. \ref{Sec:The coisotropic embedding theorem}, to perform the coisotropic embedding of $(\mathcal{M},\, \omega)$ we have to select a splitting $\mathbf{T}\mathcal{M}\,=\, K \oplus W$.
%%%
In this case a natural splitting is suggested by the decomposition of the Hilbert space $\left[\, \mathcal{H}^{\frac{3}{2}}(\mathbb{R}^3) \,\right]^3$ recalled in Sec. \ref{Sec:Electrodynamics and pre-symplectic manifolds}:
\be
\left[\, \mathcal{H}^{\frac{3}{2}}(\mathbb{R}^3) \,\right]^3 \,=\, \left[\, \mathcal{H}^{\frac{3}{2}}(\mathrm{div}0,\, \mathbb{R}^3) \,\right]^3 \oplus \mathrm{grad} \mathcal{H}^{\frac{5}{2}}(\mathbb{R}^3) \,,
\ee
whose points we denoted by $(\tilde{A},\, \psi)$.
%%%
Indeed, the second summand of the direct sum above coincide with $K$, that can be seen as the image of the following $(1,\,1)$-tensor field:
\be
P \,=\, \mathscr{G}_k \phi \otimes \frac{\delta}{\delta A_k} \,,
\ee
sometimes referred to as \textit{Coulomb connection} where:
\be
\mathscr{G}_k \,=\, \de_k \int_{\mathbb{R}^3} G_\Delta (x,\, y) \delta^{jl} \de_j \delta A_l(y) \dd^3 y \,,
\ee
$G_\Delta$ denoting the Green's function for the Laplacian.
%%%
We refer to \cite{Carinena-Ibort1985-Canonical_ghosts} for a proof that this is a projector onto the kernel of $\omega$.
%%%

\noindent On the other hand the first summand of the splitting above can be interpreted as the complement $W$ of the desired splitting.
%%%
A direct application of the coisotropic embedding theorem gives the symplectic manifold $(\tilde{\mathcal{M}},\, \tilde{\omega})$ where:
\be
\tilde{\mathcal{M}} \,=\, \mathcal{M} \times \mathrm{grad}\mathcal{H}^{\frac{5}{2}}(\mathbb{R}^3) \,,
\ee
and:
\be
\tilde{\omega} \,=\, \delta \tilde{A}_k \wedge \delta E^k + \delta \mu^k \wedge \delta \psi_k \,.
\ee
%%%

\noindent Now, the crucial point is that the vector field $\Gamma_\phi$ can be lifted to such a symplectic manifold to a vector field satisfying the hypothesis of Thm. \ref{Thm:inverse problem fode}.
%%%

\noindent Indeed, there exists a canonical way to lift a vector field from $\mathcal{M}$ to $\tilde{\mathcal{M}}$ that has been used by the author in \cite{Ciaglia-DC-Ibort-Marmo-Schiav-Zamp-2022-Symmetry} to lift the infinitesimal generator of a canonical symmetry.
%%%
In the particular situation of the example we are dealing with in the present paper the base manifold $\mathcal{M}$ is of the type $\mathcal{M} \,=\, \mathscr{H} \oplus K$ (whose points we will denote by $(m, \, k)$), where both $\mathscr{H}$ and $K$ are Hilbert spaces ($K$ representing the kernel of $\omega$). 
%%%
Thus, its tangent space at each point is isomorphic to $\mathcal{M}$ itself $\mathbf{T}_m \mathcal{M} \,\simeq \, \mathscr{H} \oplus K$ and, consequently, given the flow of a vector field $X$ on $\mathcal{M}$ preserving the splitting $\mathscr{H}\oplus K$, say:
\be
F^X_s \;\;:\;\; \mathscr{H}\oplus K \to \mathscr{H} \oplus K \;\; :\;\; (m,\, k) \mapsto ({F^X_s}_{m}(m),\, {F^X_s}_{k}(k))\,,
\ee
an action of such flow is well defined on the distribution $K$ via the tangent map:
\be
TF^X_s \;\;:\;\;  K \to K \;\; :\;\; k \mapsto  {TF^X_s}_{k}(k)\,.
\ee
%%%
The dual ${TF^X_s}^\star$ of the latter action is defined by:
\be \label{Eq: pairing}
\langle \,\mu,\, TF^X_s(k)\, \rangle \,=\, \langle \, {TF^X_s}^\star(\mu),\, k\, \rangle \,,
\ee
and, thus, a flow on the enlarged manifold $\tilde{\mathcal{M}} \,=\, \mathcal{M} \times K^\star$ is defined by:
\be
\tilde{F}^X_s \;\;:\;\; \mathscr{H} \times K \times K^\star \to \mathscr{H} \times K \times K^\star \;\; :\;\; (m,\, k,\, \mu) \mapsto ({F^X_s}_{m}(m),\, {F^X_s}_{k}(k),\, {{TF^X_{-s}}_{k}}^\star (\mu)) \,,
\ee
(the choice for the minus sign in ${{F^X_{-s}}_{k}}^\star$ is discussed in \cite{Ciaglia-DC-Ibort-Marmo-Schiav-Zamp-2022-Symmetry}). 
%%%
The lift of $X$ to $\tilde{\mathcal{M}}$ is defined to be the vector field $\tilde{X}$ whose flow is $\tilde{F}^X_s$.
%%%
In the particular case of the example considered in the present paper and considering $X$ to be the vector field $\Gamma_\phi$ defining the equations of motion, \eqref{Eq: pairing} reads:
\be
\langle \, {TF^{\Gamma_\phi}_s}_{\de \phi} (\psi_k) \, ,\, \mu^k \, \rangle \,=\, \langle \, \psi_k \, ,\, \mu^k \, \rangle  \,=\,   \langle \, \psi_k \, ,\, {{TF^{\Gamma_\phi}_s}_{\mu}}^\star (\mu^k) \, \rangle \,.
\ee
%%%
Therefore, the lift of $\Gamma_\phi$ to $\tilde{\mathcal{M}}$ reads:
\be
\tilde{\Gamma}_\phi \,=\, \left(\, \delta_{kj} E^j + \de_k \phi \, \right) \frac{\delta}{\delta A_k} + \delta^{kj} \left(\, \Delta A_j - \de_j \delta^{ml} \de_m A_l \,\right) \frac{\delta}{\delta E^k} \,.
\ee
%%%
A straightforward computation shows that:
\be
\mathcal{L}_{\tilde{\Gamma}_\phi} \tilde{\omega} \,=\, 0 \,.
\ee
%%%
Therefore, the vector field $\tilde{\Gamma}_\phi$ on the symplectic manifold $(\tilde{\mathcal{M}},\, \tilde{\omega})$ satisfies the hypothesis of Thm. \ref{Thm:inverse problem fode} and, thus, a Lagrangian description for it can be given.
%%%
In particular, there exists a (degenerate) Lagrangian $\lag \in \mathcal{F}(\mathbf{T}\tilde{\mathcal{M}})$ such that the pre-symplectic Hamiltonian system $(\mathbf{T}\tilde{\mathcal{M}},\, \tilde{\omega},\, E_\lag)$ gives, by applying the pre-symplectic constraint algorithm:
\begin{itemize}
    \item $\mathcal{M}_\infty \,=\, \tilde{\mathcal{M}}$;
    \item ${\omega_\lag}_\infty \,=\, \tilde{\omega}$;
    \item $\tilde{\Gamma}_\phi$ satisfies $i_{\tilde{\Gamma}_\phi} \tilde{\omega} \,=\, \dd {E_\lag}_\infty$.
\end{itemize}
%%%
In particular:
\be
\lag \,=\, \int_{\mathbb{R}^3} \left(\, E^k \dot{A}_k + \mu^k (\dot{\psi}_k - \de_k \phi) - E_\lag \,\right) \, \dd x^3 \,,
\ee
where $\left\{\, A_k,\, E^k,\, \mu^k,\, \dot{A}_k,\, \dot{E}^k,\, \dot{\mu}^k \,\right\}_{k=1,2,3}$ is a system of local coordinates on $\mathbf{T}\tilde{\mathcal{M}}$, $E_\lag$ is any function on $\mathbf{T}\tilde{\mathcal{M}}$ such that $\mathfrak{i}_\infty^\star E_\lag \,=\, {E_\lag}_\infty$ and ${E_\lag}_\infty$ is determined by the equation above, $i_{\tilde{\Gamma}_\phi}\tilde{\omega} \,=\, \dd {E_\lag}_\infty$, where $\tilde{\omega}$ and $\tilde{\Gamma}_\phi$ are known.
%%%
A direct computation shows that:
\be
{E_\lag}_\infty \,=\, \int_{\mathbb{R}^3} \left(\, \frac{1}{2} \delta_{jk} E^j E^k + \frac{1}{2} \delta^{jl} \delta^{km} \left( \de_j A_k - \de_k A_j \right) \left( \de_l A_m - \de_m A_l \right) \,\right) \dd x^3 \,,
\ee
giving the following Lagrangian:
\be
\lag \,=\, \int_{\mathbb{R}^3} \left(\, E^k \dot{A}_k + \mu^k (\dot{\psi}_k - \de_k \phi) - \frac{1}{2} \delta_{jk} E^j E^k - \frac{1}{2} \delta^{jl} \delta^{km} \left( \de_j A_k - \de_k A_j \right) \left( \de_l A_m - \de_m A_l \right) \,\right) \dd x^3 \,.
\ee
%%%
It is worth noting that the latter Lagrangian is equivalent to the usual Tonti-Veinberg Lagrangian for the Electromagnetic field used in Krupka's approach.
%%%

%%%%%%%%%%%%%%%%%%%%%%%%%%%%%%%%%%%%%%%%%%%%%%%%%%%%%%%%%%%%%%%%%%%%%%%%%%%%%%%%%%%%%%%%%%%%%%%%%%%%%%%%%%%%%%%%%%%%%%%%%%%%%%%%%%%%%%%%%%%%%%%%%%%%%%%%%%%%%%%%%%%%%%%%%%%%%%%%%%%%%%%%%%%%%%%%%%

\section*{Conclusions and outlines}
\label{Sec:Conclusions}
\addcontentsline{toc}{section}{\nameref{Sec:Conclusions}}

In this manuscript we argued how, starting from experimental data, i.e. curves of electric fields and vector potential, free Electrodynamics in vacuum can be naturally formulated as an infinite-dimensional dynamical system in terms of a family of vector fields over a pre-symplectic manifold that reproduce both the evolutionary part of the theory and the constraints.
%%%
We showed that, by using the coisotropic embedding theorem, the family of vector fields above can be lifted to a larger, canonical, symplectic manifold on which one can use results (Thm. \ref{Thm:inverse problem fode}) about the inverse problem available for dynamical systems described by means of a vector field over a symplectic manifold.
%%%
The details of the construction were given in Sec. \ref{Sec:The inverse problem for free Electrodynamics} where we explicitly constructed the Lagrangian.
%%%

We plan to extend this construction to the realm of non-Abelian gauge theories, having in mind the fact that they are also naturally formulated in terms of a family of vector fields over an infinite-dimensional pre-symplectic manifold.
%%%

\noindent Indeed, recall that within non-Abelian gauge theories the analogue of vector potential and electric field at each $t$ are modelled by differential $1$-forms taking values in the Lie algebra $\mathfrak{g}$ of a Lie group\footnote{Examples of physical interest include $SU(2)$, $SU(3)$ and $O(1,3)$ that are the Lie groups involved in the description of weak, strong and gravitational interactions \cite{Ibort-Spivak2017-Covariant_Hamiltonian_YangMills, Ciaglia-DC-Ibort-Mar-Schiav-Zamp2022-Non_abelian, Ciaglia-DC-Ibort-Mar-Schiav-Zamp2022-Palatini}.} (i.e. \textit{connection forms}) and vector fields taking values in its dual $\mathfrak{g}^\star$:
\be
A \,=\, A_j^a(\underline{x}) \dd x^j \otimes \xi_a \,, \qquad E \,=\, E^j_a(\underline{x}) \frac{\de}{\de x^j} \otimes \xi^a \,,
\ee
where $\left\{\, \xi_a \,\right\}_{a=1,...,\mathrm{dim}\mathfrak{g}}$ and $\left\{\, \xi^a \,\right\}_{a=1,...,\mathrm{dim}\mathfrak{g}^\star}$ are basis of $\mathfrak{g}$ and $\mathfrak{g}^\star$ respectively.
%%%
In this case $\mathcal{F} \,=\, \left[\,{\mathcal{H}^{\frac{5}{2}}(\mathbb{R}^3)}\,\right]^{3n} \times \left[\,{\mathcal{H}^{\frac{3}{2}}(\mathbb{R}^3)}\,\right]^{3n}$ (see \cite{Ciaglia-DC-Ibort-Mar-Schiav-Zamp2022-Non_abelian}), where $n \,=\, \mathrm{dim}\mathfrak{g} \,=\, \mathrm{dim}\mathfrak{g}^\star$ and solutions are curves $(A(s),\, E(s))$ over $\mathcal{F}$, with $s \in \mathbb{I} \subset \mathbb{R}$ satisfying:
\be \label{Eq:Yang-Mills}
\begin{split}
\nabla^\star_k E^k_a(s) \,&=\, 0 \;\;\; \forall \,\, s \in \mathbb{I} \,, \\
\frac{d}{ds} A_k^a(s) \,&=\, \delta_{kj} \eta^{ab} E^j_b(s) + \nabla_k \phi^a \,, \\
\frac{d}{ds} E^k_a(s) \,&=\, \delta^{kj} \eta_{ab} \left(\, -\Delta_A A_j^b(s) + \nabla_j \delta^{ml} \nabla_m A^b_l(s) + [E^k(s),\, \phi]_a \,\right) \,,
\end{split}
\ee
where $\nabla$ represents the covariant derivative associated to the connection $A$, $\nabla^\star$ is its adjoint, $\Delta_A$ is the covariant Laplacian, $\eta$ is the Cartan metric on $\mathfrak{g}$ and $\phi^a$ is an arbitrary $\mathcal{H}^{\frac{7}{2}}$ function taking values in $\mathfrak{g}$.
%%%
The above equations are \textit{Yang-Mills equations} in a space-time splitting and, as in free Electrodynamics, the first equation selects a closed  subspace of $\mathcal{F}$ \cite{Ciaglia-DC-Ibort-Mar-Schiav-Zamp2022-Non_abelian}, i.e. $\mathcal{M} \,=\, \left[\,{\mathcal{H}^{\frac{5}{2}}(\mathbb{R}^3)}\,\right]^{3n} \times \left[\,{\mathcal{H}^{\frac{3}{2}}(\nabla^\star 0,\, \mathbb{R}^3)}\,\right]^{3n}$, where $\left[\,{\mathcal{H}^{\frac{3}{2}}(\nabla^\star 0,\, \mathbb{R}^3)}\,\right]^{3n}$ is the space of $\mathcal{H}^{\frac{3}{2}}$ functions on $\mathbb{R}^3$ with zero covariant divergence. 
%%%
Again, such a subspace is a pre-symplectic manifold equipped with the structure:
\be
\omega \,=\, \delta A_k^a \wedge \delta E_a^k \,,
\ee
inherited from the cotangent bundle $\mathbf{T}^\star \left[\,{\mathcal{H}^{\frac{3}{2}}(\mathbb{R}^3)}\,\right]^{3n}$.
%%%
Again, because of the constraint equation, $\omega$ has the infinite-dimensional kernel:
\be
\mathrm{ker}\omega \,=\, <\left\{\, \nabla_k \phi^a \frac{\delta}{\delta A_k} + [E^k,\, \phi]_a \frac{\delta}{\delta E^k_a} \,\right\}> \,\simeq \, \nabla \left[\, \mathcal{H}^{\frac{7}{2}}(\mathbb{R}^3) \,\right]^n \oplus \left[ \left[\mathcal{H}^{\frac{3}{2}}(\nabla^\star 0,\, \mathbb{R}^3)\right]^{3n},\, \left[\, \mathcal{H}^{\frac{7}{2}}(\mathbb{R}^3) \,\right]^n \right] \,,
\ee
where $[\,\cdot\,,\,\cdot\,]$ denotes the Lie algebra structure of $\mathfrak{g}$.
%%%
On the other hand, the latter equations in \eqref{Eq:Yang-Mills} are, again, ordinary differential equations on the infinite-dimensional manifold $\mathcal{M}$ defining the components of a vector field.
%%%
As in the case of free Electrodynamics, such a vector field is parametrized by the arbitrary function $\phi^a \in \left[\,\mathcal{H}^{\frac{7}{2}}(\mathbb{R}^3)\,\right]^n$\footnote{This choice of regularity is made in order for its covariant derivatives to be $\mathcal{H}^{\frac{5}{2}}$ functions like the components of $A$.}, thus, the equations above define actually a whole family of vector fields on $\mathcal{M}$ that we will denote by $\Gamma_\phi$.
%%%

\noindent With this formulation in mind, we plan to carefully apply the construction of Sec. \ref{Sec:The inverse problem for free Electrodynamics} to this larger class of examples in a future work.
%%%

\section*{Acknowledgements}

The author is particularly grateful to Prof. G. Marmo, Prof. A. Ibort, Prof. A. Zampini, Dr. F. M. Ciaglia and Dr. F. Di Cosmo for the uncountable number of inspiring discussions of the last years.
%%%

\noindent The author wants also to thank an unknown referee for useful comments.
%%%

\noindent The author also acknowledges that the work has been supported by the Madrid Government (Comunidad de Madrid-Spain) under the Multiannual Agreement with UC3M in the line
of “Research Funds for Beatriz Galindo Fellowships” (C\&QIG-BG-CM-UC3M), and in the context of the V PRICIT (Regional
Programme of Research and Technological Innovation).

\bibliographystyle{alpha}
\bibliography{Biblio}

\end{document}